\documentclass[twocolumn]{aastex631}

\usepackage{amsmath}

\shorttitle{primordial scalar power spectrum with {\it Planck}}
\shortauthors{Farhang et al.}

\begin{document}

\title{Consistency of {\it Planck} Data With Power-Law Primordial Scalar Power Spectrum}
\author{Marzieh Farhang}

\author{Muhammad Sadegh Esmaeilian}
\affiliation{\scriptsize {Department of Physics, Shahid Beheshti University, 1983969411, Tehran, Iran}\\}

\begin{abstract}
In this work we explore the possibility of variations in the primordial scalar power spectrum around the power-law shape, as predicted by  single-field slow-roll inflationary scenarios.
We search for the trace of these fluctuations in a semi-blind, model-independent way in the observations of the Cosmic Microwave Background (CMB) sky.
In particular we use two sets of perturbation patterns,  specific patterns with typical features such as oscillations, bumps and transitions, as well as perturbation modes, constructed from the eigenanalysis of the forecasted or measured covariance of perturbation parameters. 
These modes, in principle, span the parameter space of all possible perturbations to the primordial spectrum, and when rank-ordered, the ones with the highest detectability would suffice to explore the constrainable features around the power-law spectrum in a data-driven (and not theoretically-biased) manner. 
With {\it Planck} measurements of CMB  anisotropies, the amplitudes of all perturbation patterns considered in this work are found to be consistent with zero. This finding confirms, in the absence of theoretical biases, the consistency of the {\it Planck} data with the assumption of power-law inflationary pattern for the primordial spectrum. 
\end{abstract}


\section{Introduction} \label{sec:intro}

The inflationary paradigm is the most widely accepted scenario to seed fluctuations in the temperature and polarization of the Cosmic Microwave Background (CMB) and the large scale distribution of matter in the Universe. The predictions of the simplest class of inflationary models, i.e., the single-field slow-roll inflation, for the primordial perturbations are Gaussian scalar and tensor fluctuations, described by power-law  spectra \citep{1979JETPL..30..682S,1982PhLB..108..389L},
\begin{equation*}
{\cal P_{\rm s}}(k)=A_{\rm s}(k/k_{\rm p,s})^{1-n_{\rm s}}, ~~~~~
{\cal P_{\rm t}}(k)=A_{\rm t}(k/k_{\rm p,t})^{n_{\rm t}} 
\end{equation*}
with $A_{\rm s,t}$ and $n_{\rm s,t}$ standing for the amplitudes and tilts of scalar (s) and tensor (t) perturbations and $k_{\rm p}$ is the corresponding pivot scale.
These predictions for the primordial scalar power spectrum  are in great agreement with CMB observations with $\ln (10^{10}A_{\rm s})=3.044 \pm 0.014$ and $n_{\rm s}=0.9649 \pm 0.0042$ at $k_{\rm p,s}=0.05 \mathrm{Mpc}^{-1}$, whereas the \%95 upper bound on the amplitude of tensor power spectrum, parametrized by the tensor-to-scalar ratio, is  $r \le 0.10$ at $k_{\rm p,t}=0.002 \mathrm{Mpc}^{-1}$ \citep{2020A&A...641A...6P,2020A&A...641A..10P}. In this work our focus is on the primordial scalar power spectrum, abbreviated as PSPS. 

Despite the great achievement, there could still be small deviations 
around these predictions. 
Various scenarios of early Universe make clear predictions for the specific patterns of the PSPS.
 For instance see
\cite{2002PhRvD..66b3511D,2003PhRvD..68f3513M,2003JCAP...05..001B,2011arXiv1104.1323C,2013PhRvD..88l3511J,2017JCAP...10..055F} for models of global oscillation
 and 
\cite{2001PhRvD..64l3514A,2007JCAP...06..023C,2011JCAP...01..030A,2012PhRvD..86f3529M,2013JCAP...10..038B} for localized oscillatory features. 
 The  parameters of the various models of early Universe have been constrained by different cosmological data \cite[e.g., ][]{2014PhRvD..89f3537M,2019PhRvR...1c3209B,2020A&A...641A..10P}.
 Forecasts have also been made on the detectability of their imprints with future surveys  \cite[e.g., ][]{2012JCAP...04..005H,2016JCAP...09..023C,2016JCAP...10..041B,2016PhRvD..94l3518X,2019PhRvR...1c3209B,2021arXiv210209007L}.

A parallel and complementary approach to this theoretically motivated path would be a model-independent analysis. In this semi-blind approach, one relaxes general degrees of freedom in the parameter space of all perturbations to the power-law PSPS, as many as numerically feasible (and required), and allow for the data to find and construct the perturbation patterns that are most tightly constrainable. 
This non-parametric search for deviation around power-law spectrum in {\it Planck} data was investigated in \cite{2020A&A...641A..10P}.
Also see \cite{2009PhRvL.103x1301Z,2011A&A...527A..49I,Farhang:2011pt,2013PhRvD..87f4026H,2014PhRvD..90b3012S,2015MNRAS.448.2232R,2016ApJ...821...30F,2017JCAP...07..042H,2018PhRvD..98d3532T,2019ApJ...871..139F,2020arXiv200401393S} 
for examples of the application of this method in different contexts in cosmology. 
In particular \cite{2021ApJ...912..104E} construct the perturbation eigenmodes to the PSPS for future CMB-S4 like and large scale surveys. 

Our goal here is to investigate the consistency of {\it Planck} observations with the power-law PSPS 
in an enhanced parameter space with different sets of degrees of freedom to cover different sorts of fluctuations. The main parameter set would be perturbation eigenmodes constructed for  {\it Planck} data. 
The major results of this work are presented in Figure~\ref{fig:traj},  showing data-driven trajectories of possible perturbations to the PSPS. As is evident from the figure, no significant deviation is detected and the PSPS is found to be consistent with the power-law spectrum. 

The organization of the paper is as follows. In Section~\ref{sec:sims} the dataset and simulations are briefly introduced. We then discuss the analysis details and introduce the patterns of perturbations explored in this work in Section~\ref{sec:anal}. The results are presented in Section~\ref{sec:res}. Section~\ref{sec:discussion} closes the paper with our final words  and a discussion of the results.

\section{simulations and data}\label{sec:sims}
In this work we use  {\it Planck} observations of fluctuations in CMB temperature and polarization \citep{2020A&A...641A...1P} to probe the physics of the early Universe through its impact on the CMB power spectra. 
 In parts we also use simulations of the CMB power spectra, generated by the publicly available Boltzmann code CAMB\footnote{https://camb.info} and {\it Planck} noise\footnote{https://pla.esac.esa.int} for comparison with results from real data, as will be discussed in Section~\ref{sec:res}.
\section{Analysis}\label{sec:anal}
We assume the underlying  PSPS of the Universe is close to the slow-roll power-law inflationary spectrum, ${\cal P}_0(k)$, with possible small deviations $\Delta {\cal P} (k)$ around it, 
\begin{equation}
{\cal P} (k)={\cal P}_0(k)+\Delta {\cal P} (k)={\cal P}_0(k)[1+\delta_{{\cal P}}(k)].
\end{equation} 
We follow the search for deviations in two different paths: search for specific, yet quite general patterns (Section~\ref{sec:spec}) and semi-blind search (Section~\ref{sec:blind}).  
The goal is then to measure the amplitudes and  other free parameters of these patterns, as will be discussed below. For parameter estimation we sample the parameter space  using the Cosmological Monte Carlo code, CosmoMC\footnote{https://cosmologist.info/cosmomc/readme.html}\citep{Lewis:2002ah}. 
This space consists of parameters characterizing perturbations to PSPS, along with standard cosmological parameters and the experimental nuisance parameters.
In the following $k_{\min}$ and $k_{\max}$ are the minimum and maximum wavenumbers considered in the analysis and we take $(k_{\rm min},k_{\rm max})\sim(0.004,1)h/{\rm Mpc}$.

The search for  features around the primordial power-law spectrum was also done in  \cite{2020A&A...641A..10P}.
 The main models used there for the reconstruction of perturbations were cubic-splines,  a sum of several top hats, a penalized likelihood method and several specific patterns. Our method differs from their work in the reconstruction scheme based on the eigenanalysis of the covariance matrix of perturbations. 
 We also search for different specific features in the spectrum except for the oscillatory pattern, which we keep for completeness.

\subsection{Specific patterns}\label{sec:spec}
\begin{figure*}
 \begin{center}
  \includegraphics[scale=0.396]{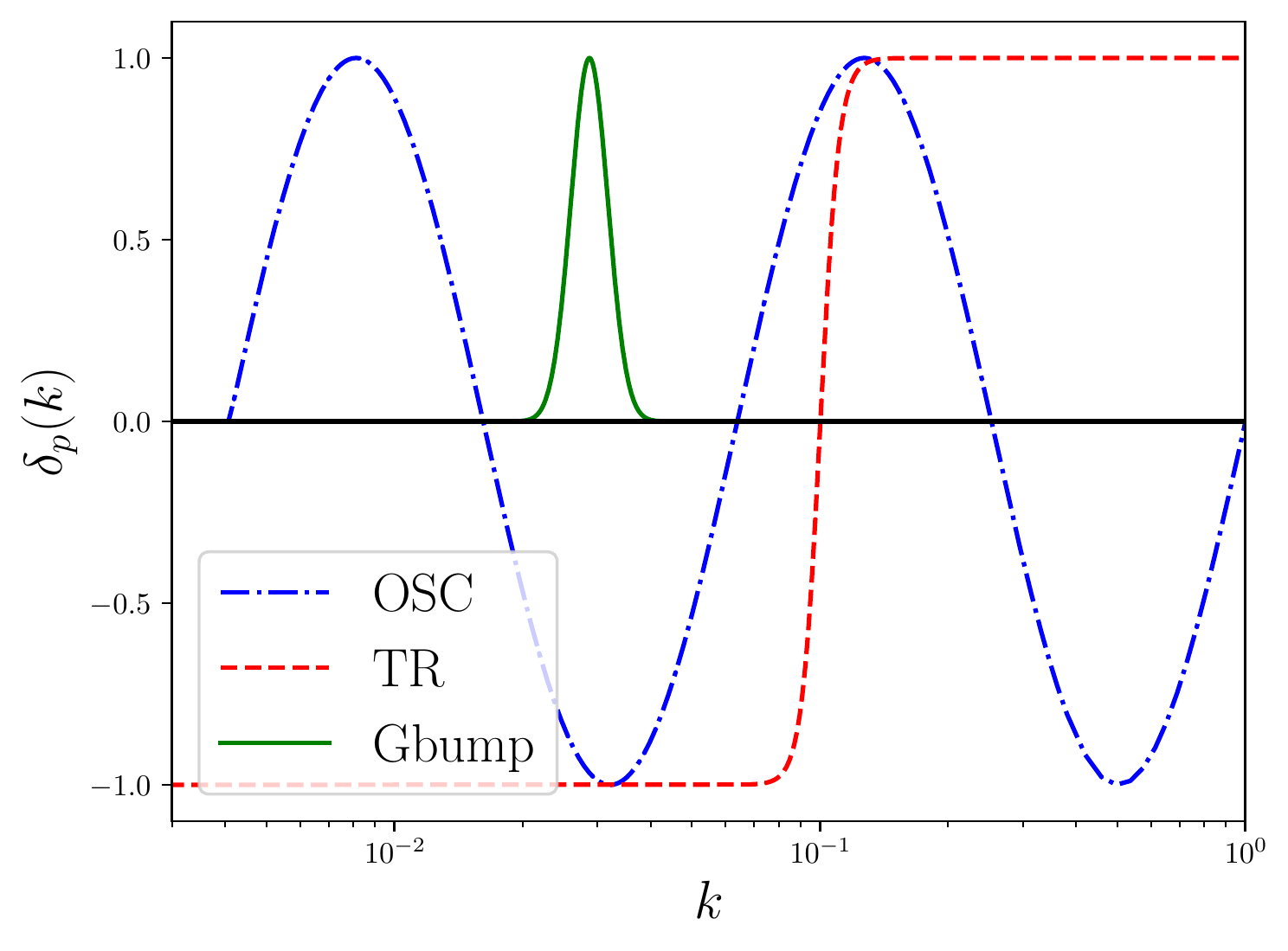}
 \includegraphics[scale=0.396]{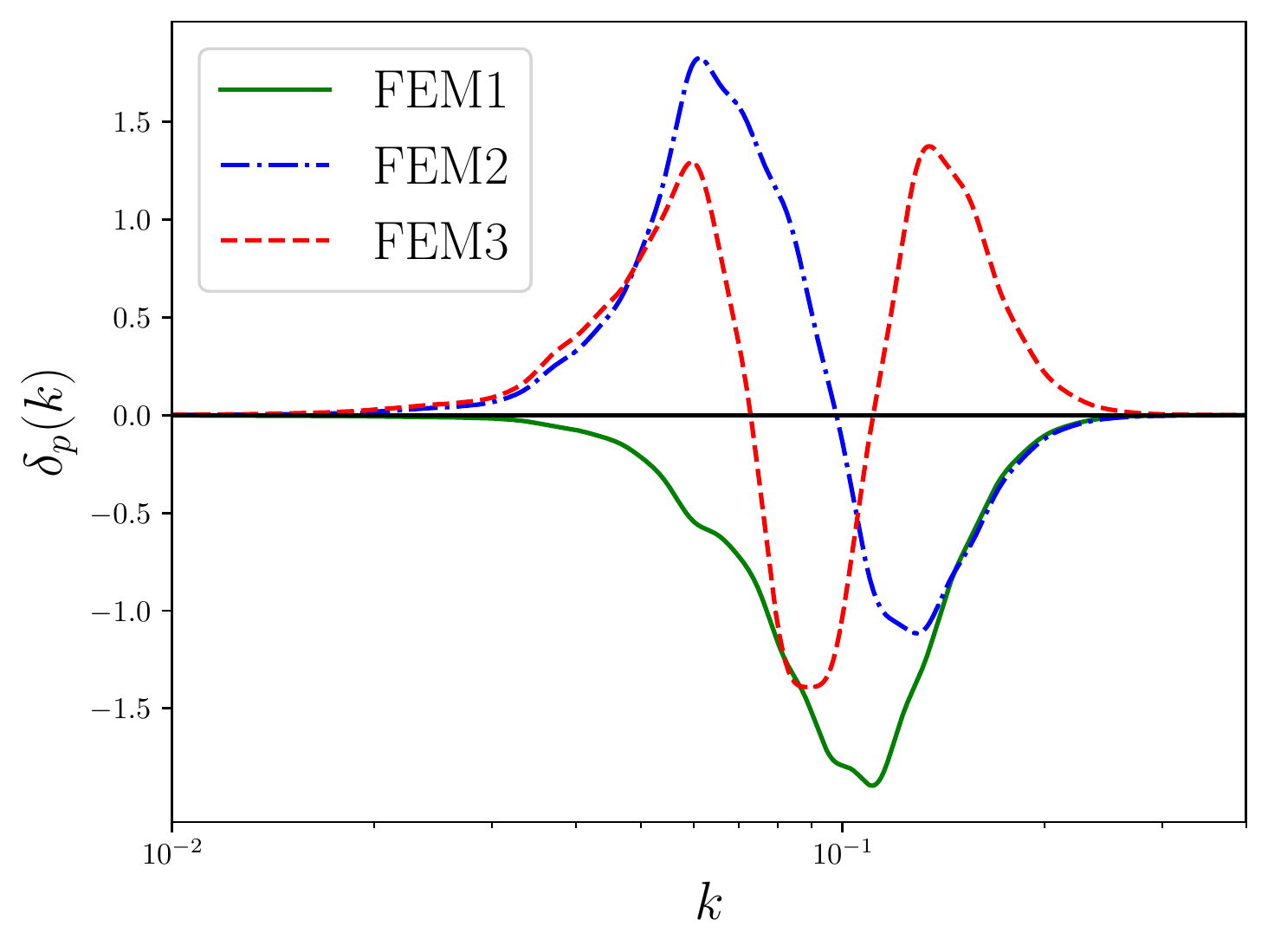}
 \includegraphics[scale=0.396]{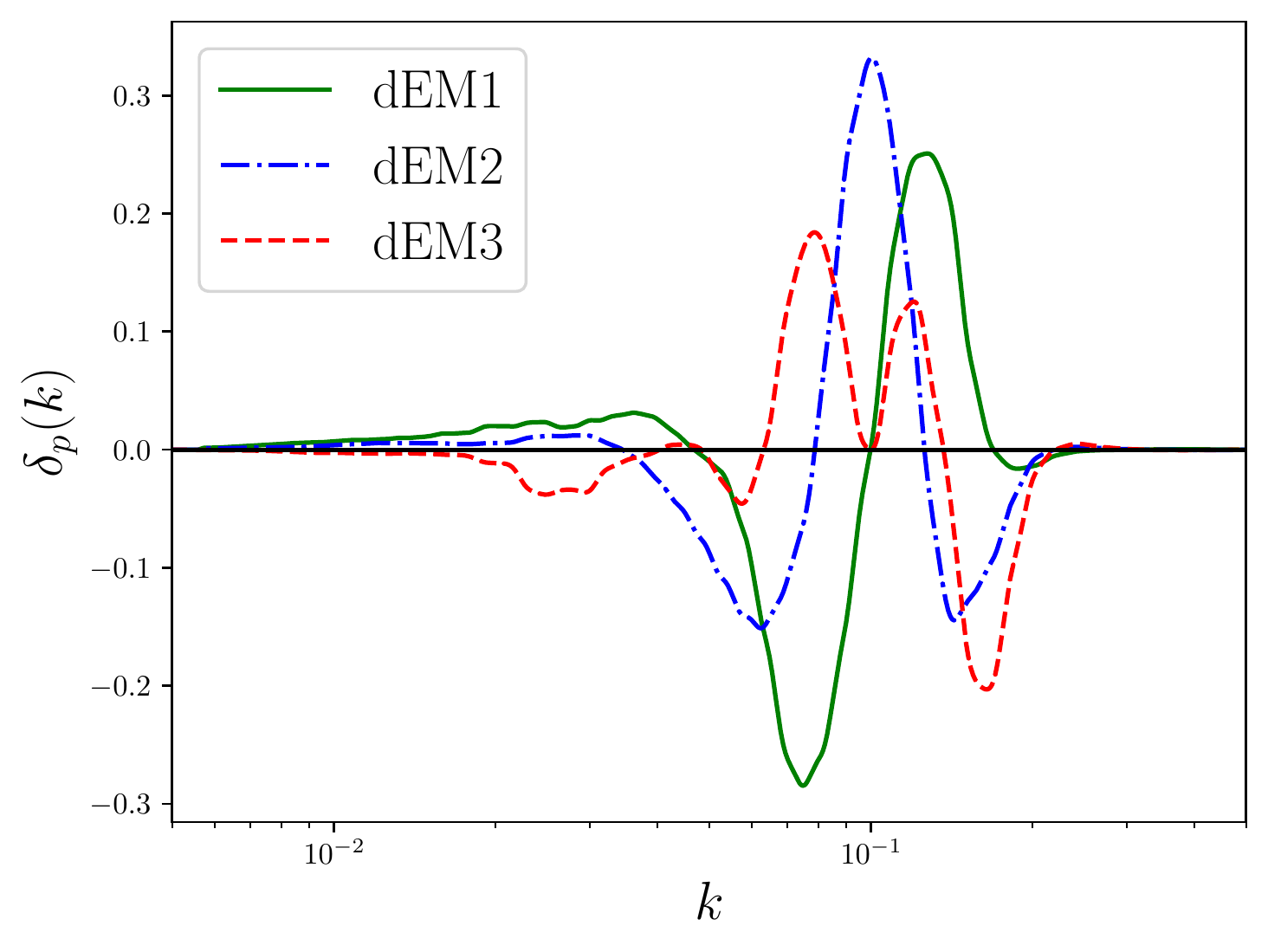}
 \includegraphics[scale=0.396]{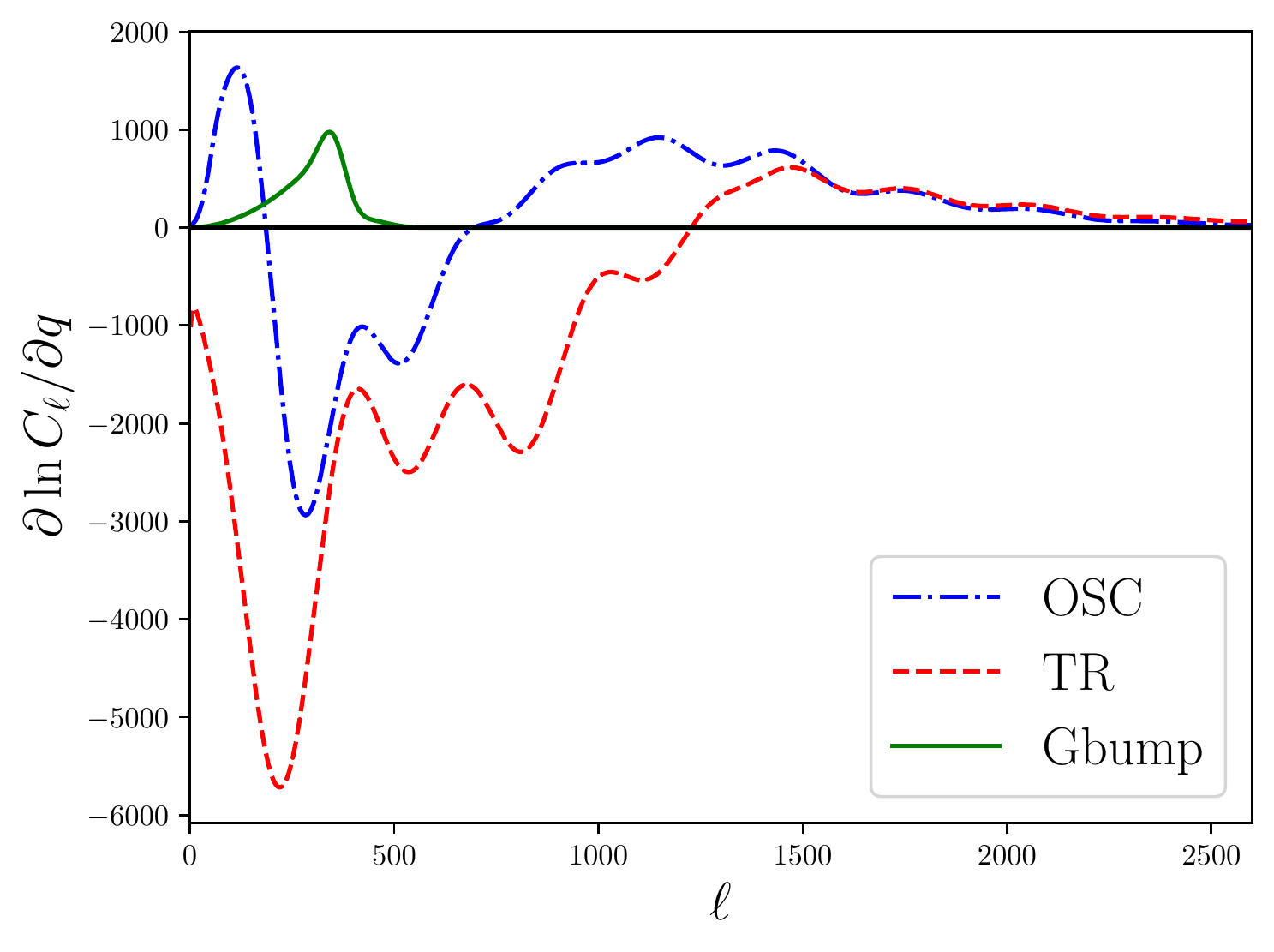}
 \includegraphics[scale=0.396]{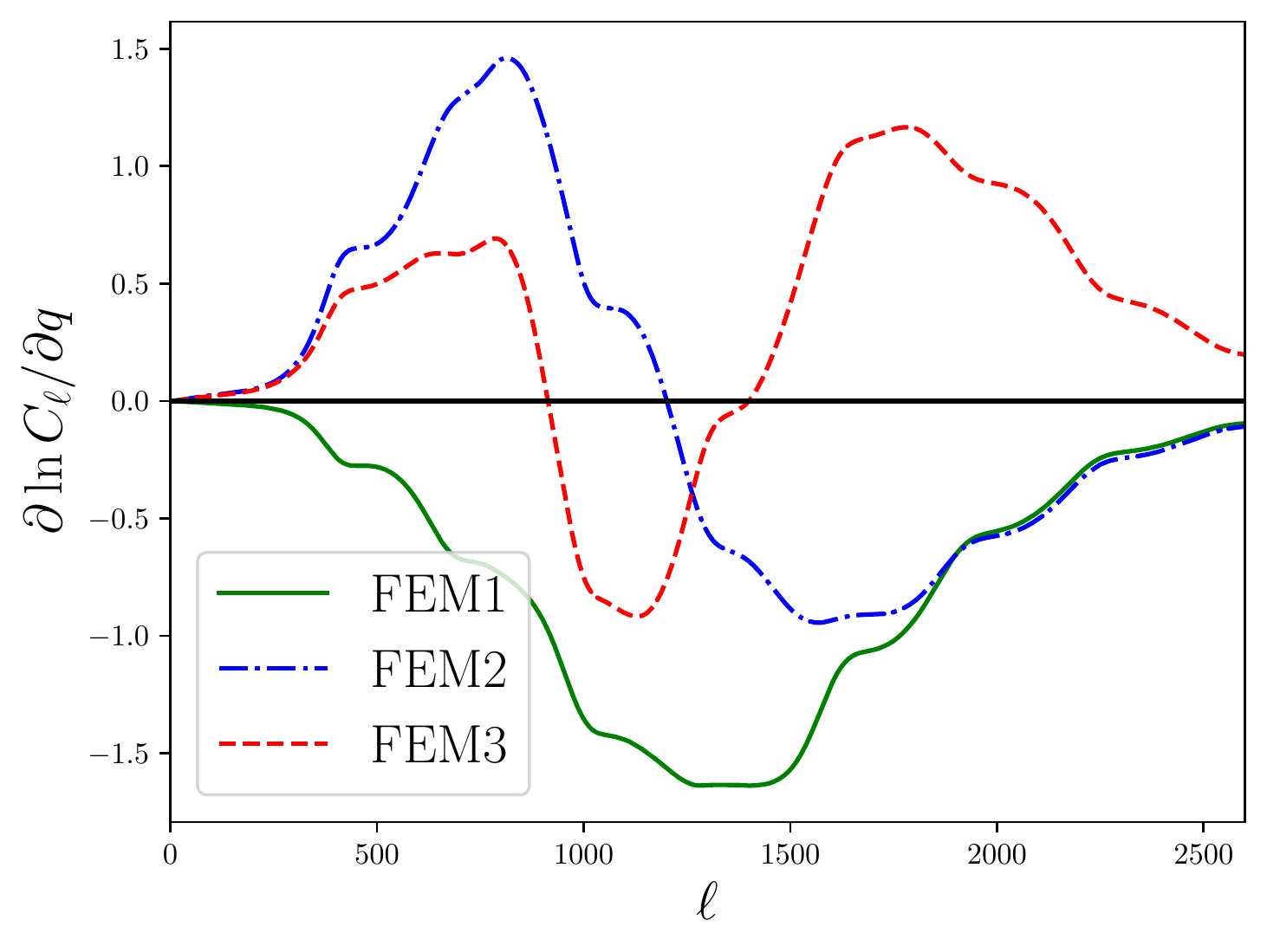}
 \includegraphics[scale=0.396]{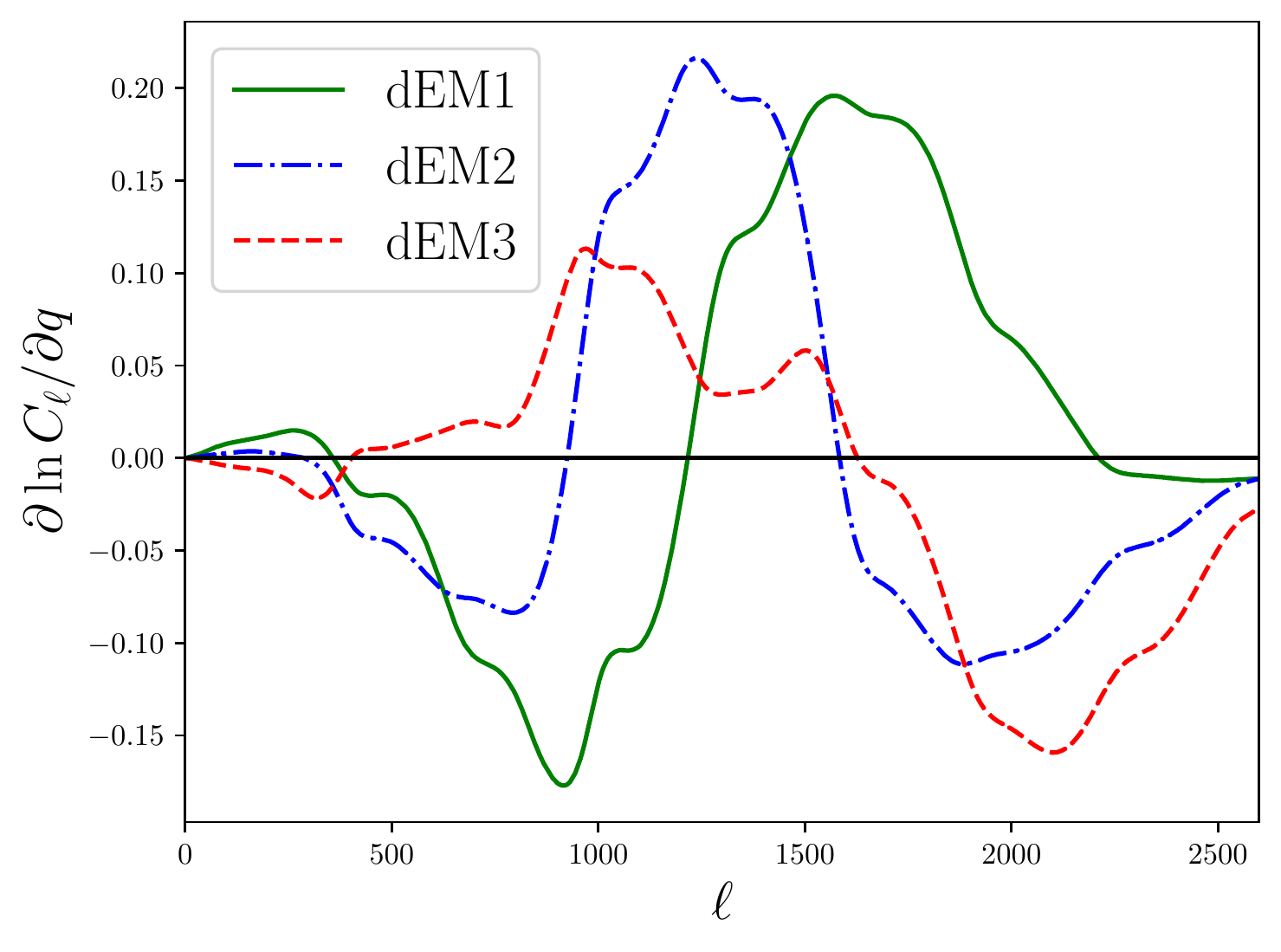}
 \caption{Top: Deviations from the power-law inflationary prediction in the form of three specific patterns: an oscillation or OSC, a transition or TR, and a Gbump (left). Three Fisher-based eigenmodes or FEMs (middle) and three data-driven eigenmodes or dEMs (right). Bottom: Response in the CMB temperature power spectrum due to small changes in the amplitude of the above perturbations. }
 \label{fig:sig}
 \end{center}
 \end{figure*}

Here we introduce the three  patterns of perturbations used in this work as generic possible forms of 
deviations around the power-law inflationary scalar power spectrum.
The left panel in Figure~\ref{fig:sig} shows these patterns (up) and the response in the CMB temperature power spectrum (bottom) to small changes in their amplitudes. \\ \\
{\bf Single Gaussian bump}, referred to as Gbump, representing a local transient feature in $k$-space,
\begin{equation}\label{eq:Gbump}
\delta_{{\cal P}}(k)=A \exp[-(\ln k -\ln k_{\rm c})^2/2\sigma^2]
\end{equation} 
where $A$, $k_{\rm c}$ and $\sigma$ are the amplitude, center and width of the bump.\\ \\
{\bf Transition} in $k$-space, modeled by a $\tanh$,
\begin{equation}
\delta_{{\cal P}}(k)=A \tanh[\alpha (\ln k -\ln k_{\rm tr})]
\end{equation} 
where $A$, $k_{\rm tr}$ and $\alpha$ are the transition amplitude, wavenumber and width  inverse. \\ \\
{\bf Oscillations} to model non-local patterns,  extended in $k$-space,
\begin{equation}
\delta_{{\cal P}}(k)=A \sin(2\pi n y)
\end{equation} 
where $A$ and $n$ (positive integer) are the amplitude and frequency of the oscillation and $y=(\ln k -\ln k_{\rm min})/\Delta \ln k$ and $\Delta \ln k=\ln (k_{\rm max}/k_{\rm min})$.

\subsection{Semi-blind patterns}\label{sec:blind}
Alongside the search for the several specific patterns in the primordial spectrum, we also investigate 
the possibility of unknown deviations from the power-law PSPS, not properly expressible by the above patterns. 
Any general deviations can in principle be expanded using a complete set of base functions that span the full $k$-range of interest \citep[see, e.g.,][for more details]{Farhang:2011pt}. We use three  sets with different motivations as discussed below.\\ \\
{\bf Gaussian bumps (Gbumps)} As the first approximation to the expansion base functions, we use $N$ Gbumps (Eq.~\ref{eq:Gbump}). 
The bump centres are logarithmically spaced in the $k$-range,  
and their widths are taken to be $\sigma=\delta \ln k /3$ with $\delta \ln k=(\ln k_{\rm max}-\ln k_{\rm min})/N$. 
We refer to this case as the multi-Gaussian expansion. 
The Gbumps can be considered as naive, however tame, approximations to Dirac delta functions.
We treat these Gbumps, in the limit of very large $n$,  as the base functions of the $N$ dimensional parameter space of all possible deviations to the scalar primordial power spectrum. 
Nevertheless, they have their numerical limitations.
 For instance, to cover all points in the $k$-space,
the widths need to be chosen so that there is non-vanishing overlap between neighbouring Gbumps . This overlap, on the other hand, destroys the orthogonality of the bumps
and can result in the correlation of the final eigenmodes constructed from the Gbumps. One could in principle correct for this error. However, as will be discussed in Section~\ref{sec:discussion}, this high precision is not required here. \\ \\
{\bf Data-driven Eigenmodes of the covariance matrix  (dEMs)} The large number of Gbumps, and the correlations between their amplitudes, in particular for neighbours, are expected to lead to relatively high uncertainties in the measurements.
 These large errors would render possible deviations hard to detect. We therefore go one step further and construct linear combinations of Gbumps with vanishing linear correlations. We call these combinations eigenmodes, or EMs, and rank order them based on their estimated uncertainties. 
We then keep only the EMs with the lowest errors and the rest are discarded. 

For the EM construction, we first generate the correlation matrix, $\bf{C}$, of the amplitudes of the $N$ Gbumps through post processing the CosmoMC results.
The goal is to linearly transform the expansion basis, here the $n$ Gbumps, so that the representation of $\bf{C}$ be diagonal in the new basis.  
These new basis functions are constructed from the eigenvectors of $\bf{C}$,
\begin{equation}\label{eq:EM}
{\cal E}_i(k)=\sum_{\alpha=1}^{N}X_{i \alpha} g_\alpha(k), ~~~~ k=1, ..., N
\end{equation} 
where $\bf{X}$ is the matrix whose columns are the eigenvectors of $\bf{C}$, $g_\alpha(k)$ refers to the Gbumps and ${\cal E}_i(k)$ is any of the $N$ perturbation eigenmodes. 
The diagonality of the covariance matrix in the new basis and the (close-to)orthogonality of the Gbumps guarantee the linear uncorrelation of the eigenmodes.
Any function representing possible deviations in the power spectrum can be expanded in terms of this new basis. The expansion coefficients yield a measure of the contribution of each mode to the perturbation.  
Moreover, the eigenvalues of $\bf{C}$ represent the (squares of the) estimated errors of the eigenmodes as measured by the data in hand.
The modes and the $C_\ell$ response to changes in the PSPS in the form of these modes are shown in the right panel of Figure~\ref{fig:sig}.
\\ \\
{\bf Fisher-based Eigenmodes (FEMs)} For comparison, we also use eigenmodes of perturbations  to PSPS constructed from the Fisher matrix. In our Fisher matrix $\mathcal{F}(\vec{q})$, the parameter set $\vec{q}$ are the amplitudes of the Gbumps and the simulations for {\it Planck} power spectrum represent the data $d$,
  \begin{equation}
  [\mathcal{F}(\vec{q})]_{\alpha\beta}=-\left< \frac{\partial^{2} \ln P_{\rm f}}{\partial q_{\alpha} \partial q_{\beta}} \right>
  \label{fish1}
  \end{equation}
 where $P_{\rm f}\equiv P_{\rm f}(\vec{q}|d)$ is the Bayesian posterior distribution of the parameter set $\vec{q}$ for the dataset $d$ and $\left<...\right>$ is the ensemble average.
 In the limit of Gaussian distribution for the parameters, one has $\mathcal{F}^{-1}={\bf C}$ and therefore the two matrices share eigenvectors. The uncorrelated modes of perturbations can thus be constructed from  Fisher eigenvectors. The details of this approach is described in \cite{2021ApJ...912..104E}.
 The only important difference is that the eigenmodes here are marginalized over the standard cosmological parameters, while the eigenmodes in  \cite{2021ApJ...912..104E} were constructed 
 with fixed standard parameters. See \cite{Farhang:2011pt} for a detailed description on marginalized mode construction. 
 
 The modes from this approach are expected to differ (although not hugely) from the dEMs in several ways. 
 The covariance matrix in the latter case was marginalized on numerous nuisance parameters (as well as standard parameters) while in the former the marginalization was only performed on the standard  parameters. Moreover, the dEM covariance matrix was based on the sampling of the parameter space while for the FEMs the assumption of the Gaussianity of the ${\bf C}$ was assumed. The Gaussianity assumption of the covariance matrix should be treated with care as the likelihood surface for the many correlated, poorly constrainable amplitudes of the Gbumps is probably far from a perfect Gaussian.
 
\section{results}\label{sec:res}
%
  \begin{table}
  \centering  
  \begin{tabular}{ccc}
  \noalign{\smallskip}
  \noalign{\smallskip}
  $A_{\rm Gbump}$&$A_{\rm TR}$ &  $A_{\rm OSC}$  \\
  \noalign{\smallskip}
  \hline 
  \noalign{\smallskip}
   $-0.079\pm  0.463$ & $0.212\pm  0.341$ & $0.009\pm 0.284$ \\
  \noalign{\smallskip}
  \end{tabular}
  \caption{The best-fit measurement of the  amplitudes of perturbations to the PSPS in the form of a single Gaussian bump (Gbump), a transition (TR) and an oscillatory pattern (OSC). The estimated $1\sigma$ uncertainty is marginalized over all other parameters included in the analysis, including the bump width and position in the Gbump, the width inverse and wavelength of the transition in the TR case and the frequency in the OSC scenario.}
  \label{tab:base}
  \end{table}
  %

%
  \begin{table}
  \centering  
  \begin{tabular}{cccc}
  \noalign{\smallskip}
  \noalign{\smallskip}
  &$A_1$&$A_2$ &  $A_3$  \\
  \noalign{\smallskip}
  \hline 
  FEM&$0.000 \pm0.002 $ & $-0.001 \pm   0.002$ & $-0.006\pm  0.004$ \\
  \hline
  dEM&$-0.002 \pm  0.016 $ & $-0.008\pm  0.016$ & $0.039\pm  0.052$ \\
  \noalign{\smallskip}
  \end{tabular}
  \caption{The best-fit measurement of the  amplitude of perturbations to the PSPS in the form of Fisher-based eigenmodes (FEMs) and data-driven eigenmodes (dEMs).}
  \label{tab:pc}
  \end{table}

We use {\it Planck} data to measure the free parameters of the perturbation patterns, introduced in Sections~\ref{sec:spec} and~\ref{sec:blind}.
 The results are presented in Table~\ref{tab:base}. As is evident from the table,  the measured amplitudes for the three specific patterns are consistent with zero. 
 For the blind search we use $60$ Gbumps with their $60$ amplitudes as the free parameters, along with the standard cosmological and observational nuisance parameters. 
All the amplitudes are found consistent with zero, as expected, since we do not see any hint (Table~\ref{tab:base}) for a deviating bump in the PSPS in the single Gbump scenario with varying $k_{\rm c}$.   
 From these $60$ Gbumps, we construct the eigenmodes dEMs, i.e., the  ${\cal E}_i(k)$'s in Eq.~\ref{eq:EM}. The measured amplitudes of these eigenmodes are presented in Table~\ref{tab:pc}, and are compared to the measurements of the Fisher-based eigenmodes. The analysis is performed with the first three modes in both cases, and no deviation from the power-law PSPS is observed. 
The result of the analysis with four eigenmodes was also null.
It is interesting to note the huge gap between the estimated uncertainties of the amplitudes for the two sets of eigenmodes (Table~\ref{tab:pc}) and the specific patterns of perturbations (Table~\ref{tab:base}).

\begin{figure}
 \begin{center}
 \includegraphics[scale=0.5]{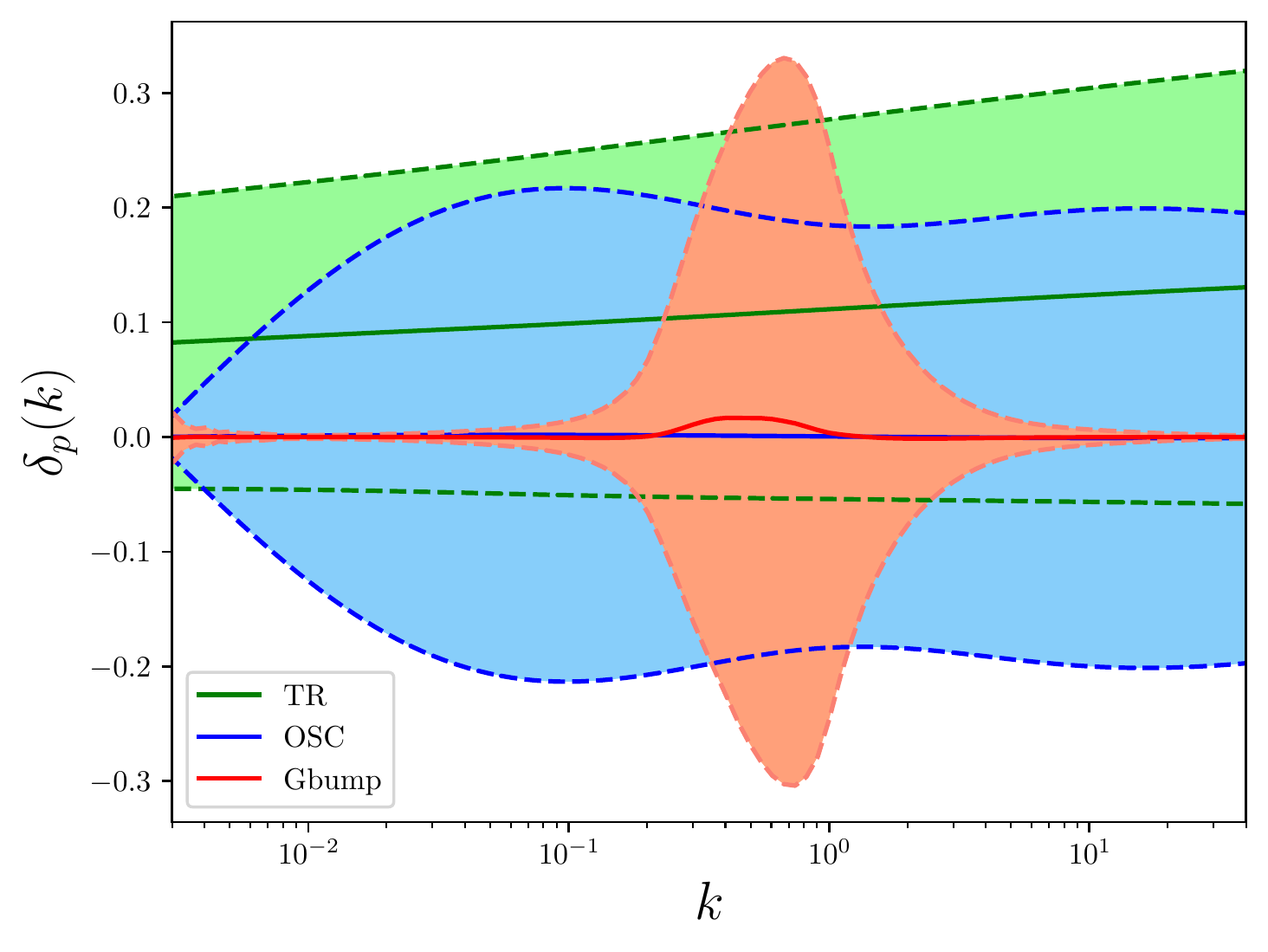}
 \includegraphics[scale=0.5]{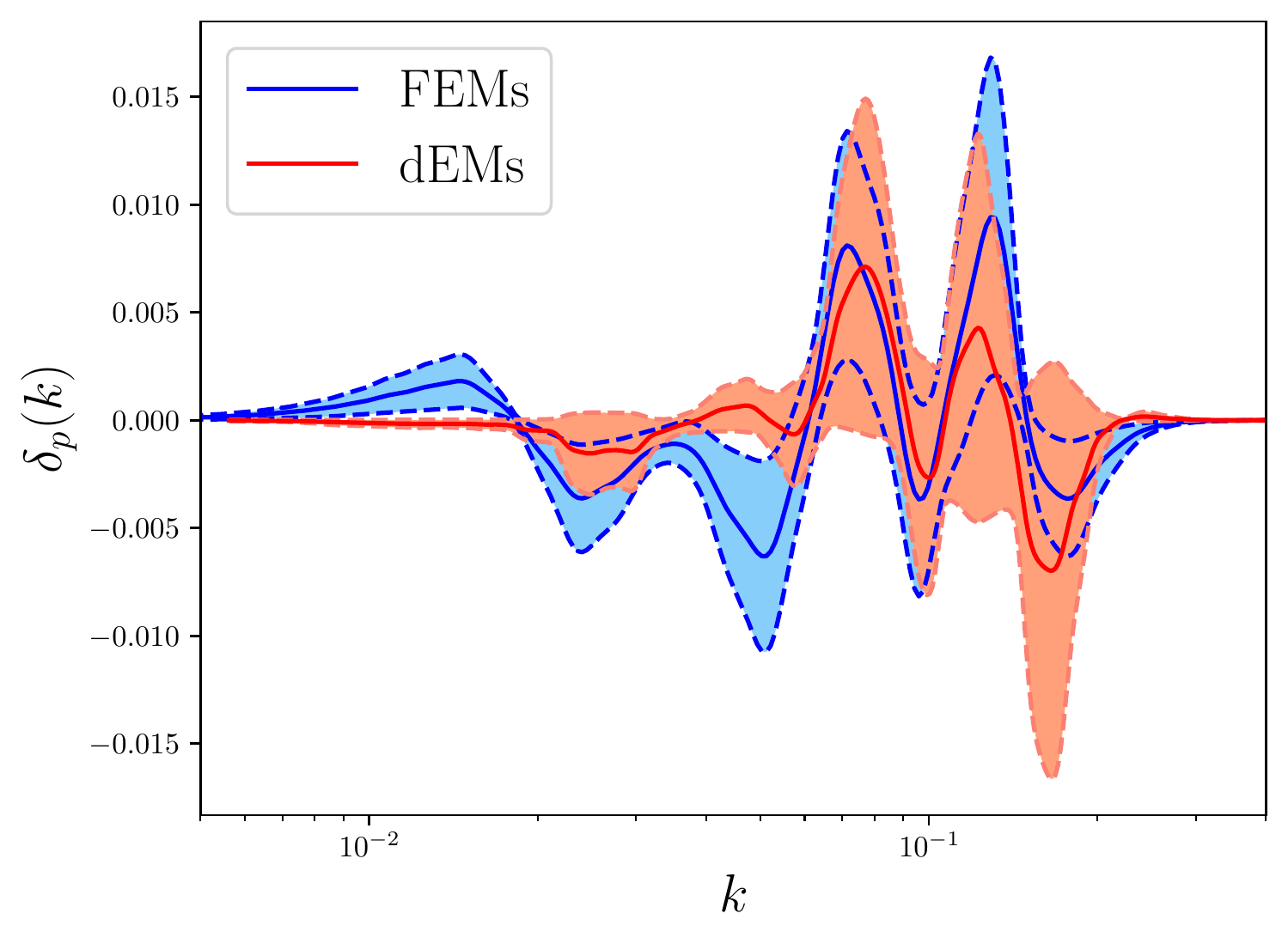}
 \caption{The reconstructed perturbations to the scalar power spectrum and the trajectories spanning the $1\sigma$ region around them, for the Gbump, oscillatory pattern and transition, labeled respectively as Gbump, OSC and TR (top), and the first three FEMs and dEMs (bottom).}
 \label{fig:traj}
 \end{center}
 \end{figure}

Figure~\ref{fig:traj} illustrates the  $1\sigma$ trajectories in the $k$-space of perturbations to the PSPS spanned by the various values taken by the parameters in each scenario.   
The reconstructed perturbations in all scenarios are consistent with power-law PSPS. It should be noted that these results are not motivated by theoretical models and are intended to be unbiased by predictions of  early Universe scenarios.

\section{discussion}\label{sec:discussion}
In this work we investigated, in a very general sense, the consistency of  CMB data, as observed by {\it Planck}, with the predictions of the slow-roll single field inflationary models.
 Specifically, we explored whether there are hints for deviations from the power-law spectrum of primordial scalar fluctuations. 
We first searched for certain, yet general, features in the spectrum, such as a Gaussian bump (with varying position and width), a transition (with varying transition wavelength and width) and an oscillatory pattern (with varying frequency). 
In parallel we also constructed orthogonal basis functions for the parameter space of all perturbations to the PSPS (up to a certain resolution, determined by the number of used basis functions) and rank-ordered them based on their  errors. 
This mode construction was done both with sampling the parameter space of perturbations and the likelihood surface exploration, and from  Fisher matrix analysis. 
We then searched for deviations in terms of these eigenmodes, focusing mainly on the first few. 
We found no hints for deviations from the power-law spectrum  in any of the above approaches, and the reconstructed PSPS was found to be fully consistent with the scale invariant scenario. 
\cite{2020A&A...641A..10P} also found null results in their non-parametric search for features in the primordial power spectrum.

There are two points in order here. First,  our method was intended to be as blind as possible to any particular theoretical model of the early Universe. The Gaussian bumps, the transitionary patterns, and the oscillations, all with varying parameters, were
used as rather typical features characterizing  general functions and were not driven by theoretical biases. The eigenmode analysis was data-driven in the sense that, by construction,  most detectable  features would show up as the first few modes  
and their amplitudes would be measured in the next steps of the analysis. 
Therefore the null results indicate that data, by themselves, do not imply any fluctuations around the scale invariant spectrum. 
However, they do not rule out the possibility of detection of fluctuations with some certain patterns (different and not constructible form the ones explored here) predicted by given theoretical models.
 Nevertheless it should be noted that these possible detections are highly model dependent and theoretically biased. 
This strong prior imposed by theory requires in turn strong theoretical justification for the preference for the certain model over the zoo of many other models of the early Universe. 

The second point is the relevance of the physics of the very early (inflationary) Universe in relaxing the Hubble tension, as reported in the disagreement between the local Universe measurements of the Hubble constant ($H_{0}=67.36 \pm 0.54 \mathrm{~km} / \mathrm{s} / \mathrm{Mpc}$)   \citep{2021ApJ...908L...6R} and the inferred value from CMB data ($H_{0}=73.3 \pm 0.8 \mathrm{~km} / \mathrm{s} / \mathrm{Mpc}$) \citep{2020A&A...641A...6P}. 
For a thorough unbiased analysis, the relevant degrees of freedom for the model of the early Universe  should be opened along with the standard cosmological parameters, including $H_0$. To appropriately address the tension, the data of the local Universe should not be included in the analysis so that it can later be compared with the final (CMB-based) inference. The hope is that the extended parameter space, including the new degrees of freedom, may change the likelihood surface in a way that the local high $H_0$ value would lie in this extended parameter region allowed by CMB data. 
However, this is not what we find in our semi-blind search. 
In particular, we find $H_0= 67.4\pm 0.6$ with the first three dEMs, implying that 
in the absence of theoretical priors, CMB data do not prefer a high $H_0$ value even in an extended parameter space encompassing new degrees of freedom in the early Universe.

\section{Acknowledgement}
Part of the numerical computations of this work was carried out on the computing cluster of the Canadian Institute for Theoretical Astrophysics (CITA), University of Toronto.

\bibliography{pinc}{}
\bibliographystyle{aasjournal}



\end{document}